\documentclass[pra, twocolumn]{revtex4-2}
\usepackage{etoolbox} 

\setcitestyle{super,open={},close={}} 
\usepackage{graphicx}
\usepackage{bbold}
\usepackage[dvipsnames]{xcolor}
\usepackage{tocloft}
\newlength{\mylen} 
\renewcommand{\cftfigpresnum}{\bf{Fig.} } 
\addtocontents{toc}{\protect\thispagestyle{empty}}
\usepackage{float}
\usepackage{amsmath}
\usepackage{textcomp,gensymb} 
\usepackage{lettrine} 

\usepackage[colorlinks=true,linkcolor=blue,citecolor=blue,filecolor=blue,urlcolor=blue]{hyperref}

\usepackage[english]{babel}

\begin{document}

\title{Enhancing second-harmonic generation with electron spill-out at metallic surfaces}

\author{Muhammad Khalid}
\email{muhammad.khalid@iit.it}

\author{Cristian Cirac\`i}
\email{cristian.ciraci@iit.it}
\date{\today}
\affiliation{Istituto Italiano di Tecnologia, Center for Biomolecular Nanotechnologies, Via Barsanti 14, 73010 Arnesano, Italy.}

\renewcommand{\figurename}{\textbf{Fig.}}

\begin{abstract}
\begin{center}
\section*{Abstract}
\end{center}
Second-order nonlinear optical processes do not manifest in the bulk of centrosymmetric materials, but may occur in the angstroms-thick layer at surfaces. At such length-scales, quantum mechanical effects come into play which could be crucial for an accurate description of plasmonic systems.
In this article, we develop a theoretical model based on the quantum hydrodynamic description to study free-electron nonlinear dynamics in plasmonic systems.
Our model predicts strong resonances induced by the spill-out of electron density at the metal surface.
We show that these resonances can boost second-harmonic generation efficiency up to four orders of magnitude and can be arbitrarily tuned by controlling the electron spill-out at the metal surface with the aid of thin dielectric layers.
These results offer a possibility to artificially increase nonlinear susceptibilities by engineering optical properties at the quantum level.
\end{abstract}

\maketitle
\patchcmd{\section}
  {\centering}
  {\raggedright}
  {}
  {}
\patchcmd{\subsection}
  {\centering}
  {\raggedright}
  {}
  {}

\thispagestyle{plain} 

\section*{Introduction}\label{sec:Introduction}
\lettrine[]{M}{} odern photonic devices rely on nonlinear optical effects to carry out functionalities such as parametric tuning of the laser light spectrum, generation of ultra-short pulses, all-optical signal processing and ultrafast switching \cite{Sidick1994Opt.Lett., Krasavin2018Laser&PhotonicsReviews}.
Furthermore, researchers are constantly extending the reach of nonlinear optical phenomena to a growing variety of practical applications, such as nonlinear microscopy \cite{Boyd2006}, ultrasensitive-optical shape characterization \cite{Butet2013NanoLett.}, monitoring the processes in chemical synthesis of nanostructures \cite{Sauerbeck2014ACSNano} and nonlinear plasmonic sensing \cite{Butet2014Nanoscale}.

The inherent weakness of optical nonlinearities however has strongly hindered the chip integration of all-optical devices. 
Conventionally, to enhance the intrinsic nonlinear response of natural materials, one has to rely on phase-matching or quasi-phase matching techniques, which require relatively long propagation distances.
Distances can be shortened to several microns by employing photonic band-gap crystals, which is in general detrimental for the band width.
A different route to increase nonlinear optical response can be achieved through plasmonic effects  -- collective oscillations of conduction electrons -- whose near-field enhancement properties can effectively increase the nonlinear susceptibilities. 
Plasmons are not limited by diffraction and allow to focus the light in regions where dielectric nonlinearities can be embedded, or such that metallic nonlinearities themselves can be enhanced.
The advantage of such approach is that one can effectively increase local nonlinear susceptibilities in nanoscale devices.
The drawback, however, is that it is hard to combine building-up mechanisms that allow to increase the overall conversion efficiencies.
In order to overcome this limitation then it is necessary to artificially increase nonlinear susceptibilities by engineering optical properties at the quantum level.
At THz frequencies, for example, this can be obtained using semiconductor quantum wells which are known to offer the largest nonlinear susceptibilities \cite{Lee:2014ewa}.
Increasing the carrier density, however, causes a transition from size-quantization  to the classical regime of plasmon oscillations, in the visible range of frequencies. 


Since the pioneering works on nonlinear optics \cite{Terhune1962Phys.Rev.Lett., Brown1965Phys.Rev.Lett., Jha1965Phys.Rev., Bloembergen1966Phys.Rev.}, SHG has been continuously studied both experimentally \cite{Bloembergen1968Phys.Rev., Sonnenberg1968J.Opt.Soc.Am., Simon1974Phys.Rev.Lett., Quail1985Phys.Rev.B, Coutaz1987JournalofAppliedPhysics, Pedersen1989_JOSAB, Urbach1992Phys.Rev.B, Chang1997SurfaceScience, Krause2004JournalofAppliedPhysics, ODonnell2005NewJ.Phys., Grosse2012Phys.Rev.Lett., Galanty2018LightSciAppl}, and theoretically \cite{Rudnick1971Phys.Rev.B, Sipe1980Phys.Rev.B, Sipe1980SolidStateCommunications, Keller1985_PRB, Keller1986_PRB, Corvi1986Phys.Rev.B, Maytorena1995Phys.Rev.B, Chizmeshya1988Phys.Rev.B, Weber1987Phys.Rev.B, Liebsch1988Phys.Rev.Lett., Hu1989Phys.Rev.B, Heinz1991, Schaich2000Phys.Rev.B}.
In centrosymmetric materials, such as noble metals, second-order nonlinearities in the bulk region are intrinsically forbidden \cite{Boyd2006}, nevertheless, the symmetry can be broken at the material surface, giving rise to a surface SHG mechanism \cite{Vollmer1996SurfaceScience, Butet2015ACSNano}.
Typically, the SHG surface process occurs within a very thin layer of the order of few angstroms (a few Thomas-Fermi screening lengths) at the surface where the induced charges are confined. 
At such near-atomic length scales, classical electrodynamics fails to address the microscopic details and consideration of nonlocal and quantum mechanical effects may become crucial for accurately characterizing optical behavior of a metallic system.

In this context, the simplest tool dealing with nonlocal electron pressure is the Thomas-Fermi hydrodynamic theory (TFHT)  \cite{Ciraci2012Science, Ciraci2013ChemPhysChem, Raza2015J.Phys.:Condens.Matter, Yoo2019NatCommun, Hua1986Phys.Rev.B, Ciraci2015Phys.Rev.B, Scalora2018Phys.Rev.A}. 
The TFHT is often combined with the hard-wall boundary condition, which overlooks the essential quantum mechanical effects like electron spill-out and quantum tunneling \cite{Ciraci2012Science, Ciraci2013ChemPhysChem, Raza2015J.Phys.:Condens.Matter, Yoo2019NatCommun, Yan2015Phys.Rev.B}.
While some efforts have been made to include in the hydrodynamic description space-dependent electron density profiles, such as linear and quadratic shape functions \cite{Sipe1980Phys.Rev.B, Sipe1980SolidStateCommunications, Corvi1986Phys.Rev.B, Maytorena1995Phys.Rev.B}, the TFHT remains inadequate to deal with the realistic density profile, as it yields spurious modes originating from the exponentially decaying tail of the density \cite{Yan2015Phys.Rev.B, Ciraci2016Phys.Rev.B}. 
The significance of an accurate description of the charge density profile in SHG has already been pointed out in the low-frequency regime both using a full quantum mechanical treatment\cite{Weber1987Phys.Rev.B, Liebsch1988Phys.Rev.Lett., Hu1989Phys.Rev.B, Heinz1991, Schaich2000Phys.Rev.B} as well as an orbital-free approach\cite{Chizmeshya1988Phys.Rev.B}.

Recently, a variety of methods based on effective local parameters have been proposed to deal with electron spill-out effects.
These approaches are generally developed for the linear response regime\cite{Esteban2012NatCommun, Yan2015Phys.Rev.Lett., Kupresak2018Adv.TheorySimul.} and their generalization to the nonlinear dynamics is not trivial.
The TFHT, on the other hand, is intrinsically nonlinear and it has long been used to describe the nonlinear response of metals\cite{Hua1986Phys.Rev.B, Ciraci2015Phys.Rev.B, Scalora2018Phys.Rev.A}.
In order to overcome its limitation, one should consider $\nabla n$-dependent corrections ($n$ being the electron density) to the free-electron gas kinetic energy, namely adding the von Weizs\"acker term to the TF kinetic energy \cite{Toscano2015NatCommun, Ciraci2016Phys.Rev.B, Ciraci2017Phys.Rev.B}.
Such approach is usually referred to as quantum hydrodynamic theory (QHT).
Over the past few years, QHT has been applied to study linear optical properties of individual nanoparticles as well as of extremely coupled nanostructures (in the tunneling regime) and a very good agreement with the time-dependent density functional theory calculations has been reported \cite{Ciraci2016Phys.Rev.B, Ciraci2017Phys.Rev.B, Khalid2018Opt.Express, Khalid2019Photonics}.
The QHT can also incorporate classical nonlinear effects, such as Lorentz and convective contributions, in multiscale plasmonic systems with deep sub-wavelength features, where the quantum effects cannot be neglected.

In the present work, we expand the QHT equations beyond the linear approximation to probe second-order nonlinearities in plasmonic structures of arbitrary shapes and sizes.
Our method can efficiently handle realistic profiles of the charge density and indeed predicts three contributions in the SHG process that emerge from the spatial dependence of the charge density.
We examine the spectral dependence of SHG from Na and Ag thick films and show that the SHG efficiency spectra can exhibit a strong resonance due to the electron spill-out at the metal surface.
Our results show that quantum electron spill-out could be used to boost nonlinear efficiencies up to 4 orders of magnitude.
If successfully proven experimentally, our work could enable efficient nonlinear optics at the nanoscale.
\section*{Results}
\noindent In QHT the pressure term takes the more general form of $\mathrm{\nabla }\frac{\delta G[n]}{\delta n}$
where $G[n]$ is the energy functional of the free-electron gas:
\begin{equation}
G[n] = T_{\textrm{TF}}[n]+\lambda T_{\textrm{W}}[n, \nabla n]+E_{\textrm{xc}}[n],
\label{eq:G[n]}
\end{equation}
where $T_{\textrm{TF}}[n]$ is the TF kinetic energy functional, $T_{\textrm{W}}[n, \nabla n]$ is the gradient-dependent correction term (von Weizs\"acker term) and $E_{\textrm{xc}}[n]$ is the exchange-correlation energy functional in the local density approximation.
The parameter $\lambda$, weighting the von Weizs\"acker functional, is of extreme importance as it defines the decay of the electron density and it is usually taken as $1/9 \leq \lambda \leq 1$ \cite{Ciraci2016Phys.Rev.B}.
The nonlinear QHT expansion up to second-order terms introduces three groups of nonlinearities in addition to the classical nonlinear hydrodynamic terms \cite{Scalora2010Phys.Rev.A, Ciraci2012Phys.Rev.B} (see the Methods section for more details):
i) a  convective term, proportional to $\nabla n_0/n_0$, arising from the spatial dependence of the electron equilibrium density $n_0({\bf r})$; ii) a quantum pressure term given by the product of the linear induced density $n_1$ and the gradient of the linear quantum pressure, i.e., $ n_1\nabla \left(\frac{\delta G[n]}{\delta n} \right)_1$; iii) a second-order quantum pressure term, $\nabla \left(\frac{\delta G[n]}{\delta n} \right)^{\mathrm{NL}}_2$, that emerges from the nonlinear dependence of the kinetic functional from the electron density $n$.
The last two terms generalize the nonlinear quantum pressure term in Refs. \citenum{Scalora2010Phys.Rev.A} and \citenum{Ciraci2012Phys.Rev.B} and introduce a set of contributions that are inversely proportional to some powers of $n_0$.
Note that because $n_0\to 0$ near the metal surface, these contributions, as well as the convective contributions, are expected to be important at the surface.
Further details about the nonlinear QHT are reported in the Methods section and Supplementary Note 1. 
\begin{figure}
	\centering
	\includegraphics[width=0.95\linewidth]{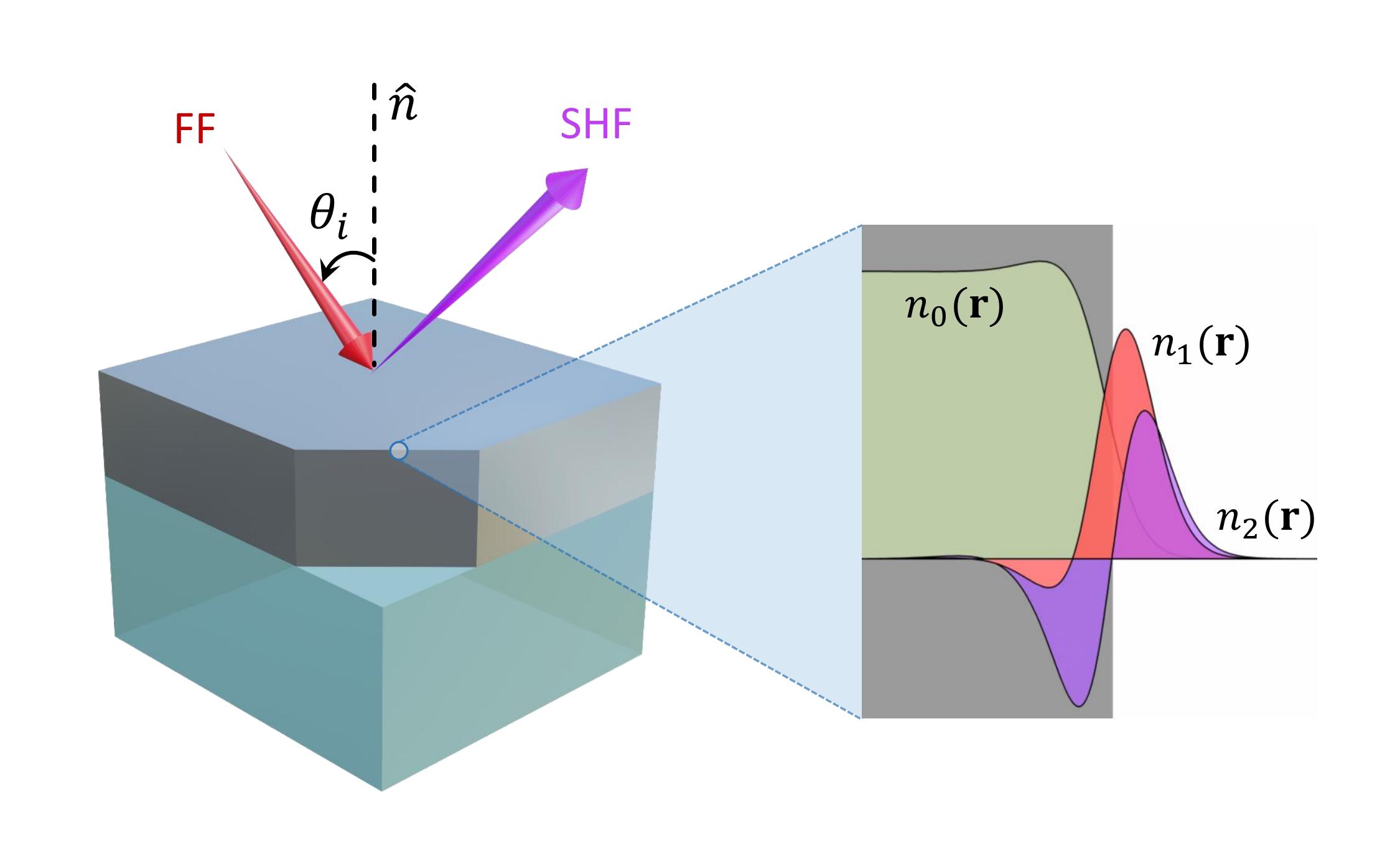}
	\caption{\textbf{Schematic representation of the numerical setup for SHG.} A thick metallic film placed on a glass substrate and illuminated by a transverse magnetic polarized plane wave making an angle $\theta_i$ with the unit normal vector $\hat{n}$. The fundamental field (FF) and the second-harmonic field (SHF) correspond to the fields of the pump and the generated signal, respectively. A zoom-in at the metal-air interface depicts spatial variation of the static and induced charges densities at the fundamental and second harmonics.}
	\label{fig:Slab_Geometery}
\end{figure}

We employ the nonlinear QHT to investigate the linear and second-order nonlinear optical properties of Na and Ag films, illuminated by a fundamental field with energy $E=\hbar \omega_1$ impinging at an angle $\theta_i$ with respect to the normal unit vector $\hat{n}$ at the metal surface, as depicted in Fig.~\ref{fig:Slab_Geometery}. We assume that the electric field is transverse magnetic (TM) polarized, i.e., laying in the incidence plane, as a transverse electric (TE) polarized wave would not induce any accumulation of charge at the metal interface. To make the system more realistic, we place the metal film on a glass substrate, although its impact on SHG is negligible for sufficiently thick films.

The SHG efficiency at $\omega_2=2\omega_1$ can be defined as:
\begin{equation}
\text{SHG efficiency}=\frac{P_{\rm SH}(\omega_2)}{P_{\rm FF}(\omega_1)} 
\end{equation}
where $P_{\rm SH}(\omega_2)$ is second-harmonic reflected power from the metal interface and $P_{\rm FF}(\omega_1)$ represents the input power of the exciting (fundamental) field. In the following, we consider the pump intensity of $I_0=80$~MW/cm$^2$.
\subsection*{A simple Drude-like metal}\label{subsec:NaFilm}
\begin{figure*}
	\centering
	\includegraphics[width=0.95\textwidth]{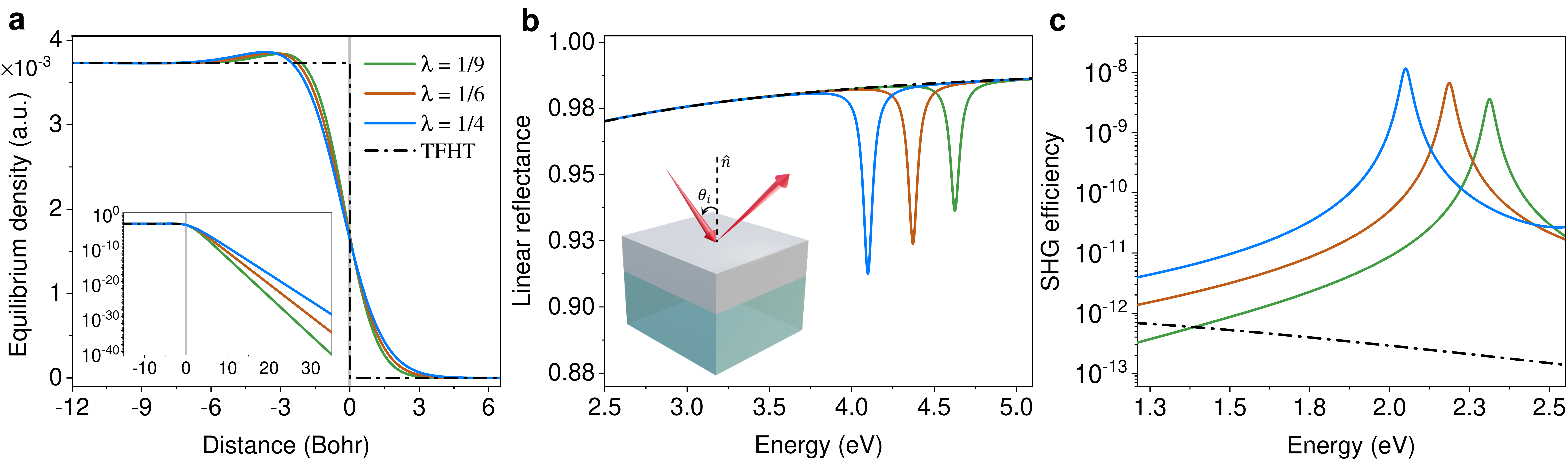}
	\caption{\textbf{Linear and nonlinear response of a 400 nm thick Na film under transverse magnetic field excitation.} (\textbf{a}) Equilibrium electron density calculated self-consistently for different values of spill-out parameter $\lambda$. For Thomas-Fermi hydrodynamic theory (TFHT) a constant density inside the metal with no spatial dependence is assumed. The inset shows the decay of the electron density for each $\lambda$ and the solid grey line indicates the position of the metal interface. (\textbf{b}) Linear reflectance and (\textbf{c}) SHG efficiency as a function of exciting (fundamental) energy of the incident field at $\theta_i=75\degree$ for different types of charge density profiles. We consider the peak intensity of the incident field $I_0=80$~MW/cm$^2$ and the electron damping rate $\hbar\gamma=0.066$~eV.}
	\label{fig:NaResults}
\end{figure*}

\noindent Let us first consider a simple Drude-like metal system, a Na film placed in free space.
Optical response of Na is easier to understand as the effect of interband transitions can be neglected at visible frequencies. 
We numerically analyze the spectral dependence of a 400 nm thick Na film excited by a TM-polarized plane wave.
The self-consistent equilibrium charge density, calculated using equation (\ref{eq:scn0}) for different values of the spill-out parameter $\lambda$ (see the Methods section), is shown in Fig.~\ref{fig:NaResults}a.
$\lambda=1/9$ is associated to a smaller spill-out and a faster decay of the electron density whereas $\lambda=1/4$ represents a larger spill-out and a slower decay, as can be seen in the inset of Fig.~\ref{fig:NaResults}a.
The equilibrium density in the case of TF approximation is also plotted. In this case, electrons are not permitted to spill from the metal surface into the free space (hard-wall boundary condition).
Since the impact of interband transitions in Na can be neglected, we set the core dielectric constant, $\varepsilon_\infty=1$.
In Fig.~\ref{fig:NaResults}b, we plot the linear reflection spectra. As expected, because of the absence of any structuring on the Na film surface, the TFHT does not present any particular feature.
On the other hand, the spatial dependence of the charge density in the QHT produces a dip in the reflection spectra that depends on the equilibrium density decay rate at the metal-air interface.
As the electron spill-out increases (for larger values of $\lambda$), the resonance dip moves to lower energies.
This means that, in principle, by controlling the amount of electron spill-out from the metal surface, one could tune the resonance to the frequency of interest.
These resonances are due to excitation of multipolar surface plasmons and are a consequence of electron spill-out exclusively.
In the TF approximation (as well as in the standard local case not reported here), in fact, no resonant behavior can be observed.

The existence of this multipolar collective mode was predicted several decades ago for sufficiently diffused electron density profiles \cite{Bennett1970Phys.Rev.B} and was also observed directly in the electron energy loss spectroscopy (EELS) experiments \cite{Tsuei1990_PRL}. This mode is expected to appear between the ordinary (Ritchie’s) surface plasmon and bulk plasmon modes. Experiments performed on a variety of simple metals for the investigations of multipole plasmons accord well with the theoretical predictions \cite{Liebsch1987_PRB, Liebsch1991_PRL, Tsuei1991_SurfSci, Sprunger1992_SurfSci}.
However, contrarily to the free-electron metals, there has been a controversy on the existence of the multipole mode in noble metals (i.e., Ag), which show complicated behavior due to filled 4d bands. EELS experiments on Ag does not show any clear evidence of the multipole plasmons. Nevertheless, experiments performed on Ag using high-resolution energy-loss spectroscopy low-energy electron diffraction revealed a signature between the surface and bulk plasmon modes and it was attributed to the excitation of multipole plasmons \cite{Moresco1996_PRB}. Later on, Liebsch \cite{Liebsch1998Phys.Rev.B} argued that the interpretation of the multipole mode in Ref. \citenum{Moresco1996_PRB} might not be correct and the multipole mode in Ag, instead of appearing before, should exist after the bulk plasma frequency. The existence of Ag multipole plasmon was also questioned due to the fact that the surface and bulk plasmons lie in a close proximity. In quest of the multipolar plasmon in noble metals, experiments were performed on Ag using a more suitable and sophisticated experimental technique, namely the angle- and energy-resolved photoyield, which confirmed the existing of the multiple plasmon mode both in the Ag bulk single-crystal surfaces and its adlayers \cite{Barman2004_PRB, Barman200_SurfSci}. The mode predicted by Liebsch \cite{Liebsch1998Phys.Rev.B}, however, was not observed in these experiments. On the other hand, in Ref. \citenum{Politano2013_Plasmonics} and \citenum{Politano2013_JPCM}, the authors performed experiments on thin Ag films grown on Ni and showed that they found a mode in agreement with the Liebsch's prediction, however, this mode can only be excited under certain experimental conditions (i.e., enhanced surface sensitivity).
The reader may refer to Ref. \citenum{Pitarke2007Rep.Prog.Phys.} and the references therein for a detailed overview of the theoretical and experimental studies of higher-order multipolar surface plasmons.

These spill-out induced resonances, as shown in Fig.~\ref{fig:NaResults}b, can be exploited to enhance the SHG efficiency by several order of magnitude (see Fig.~\ref{fig:NaResults}c).
The nonlinear QHT predicts, in fact, a very large enhancement of the SHG efficiency when the generated frequency matches the multipolar resonance in the linear spectrum.
This kind of peak in the SHG intensity was also predicted in the nonlinear analysis of potassium surface within the TFHT \cite{Maytorena1995Phys.Rev.B}.
Although the authors in this work used unrealistic equilibrium density, nonetheless, they pointed out that the position and strength of this peak strongly depend on the shape of the charge density profile.
Our QHT calculations show, as depicted in Fig.~\ref{fig:NaResults}c, that SHG enhancement can exceed 4 orders of magnitude than the conversion efficiency away from resonance. 
\subsection*{Fitting experimental data of SHG from Ag films }\label{subsec:AgFilm}
\noindent Simple Drude-like metals, such as Na, are convenient from a theoretical point of view since they present an ideal optical behavior. In practice, however, they are very hard to work with because of their high reactivity. 
It would be interesting to analyze the possibility to observe a similar phenomenon in commonly used metals in plasmonics, such as noble metals.
Noble metals' optical response is complicated by the contribution of the interband transitions, which can act on many levels. They modify the linear susceptibility, contribute to the nonlinear response \cite{Scalora2014Phys.Rev.A} and affect both the equilibrium charge density and the induced charge distribution at the metal surface \cite{Toscano2015NatCommun, Teperik2013Phys.Rev.Lett., Liebsch1998Phys.Rev.B}.
In the following, we neglect the nonlinear contribution from interband transitions and account for all the other effects by considering a polarizable medium of $\varepsilon_\infty > 1$ permeating the whole metal volume. 

\begin{figure}
\centering
\includegraphics[width=0.8\linewidth]{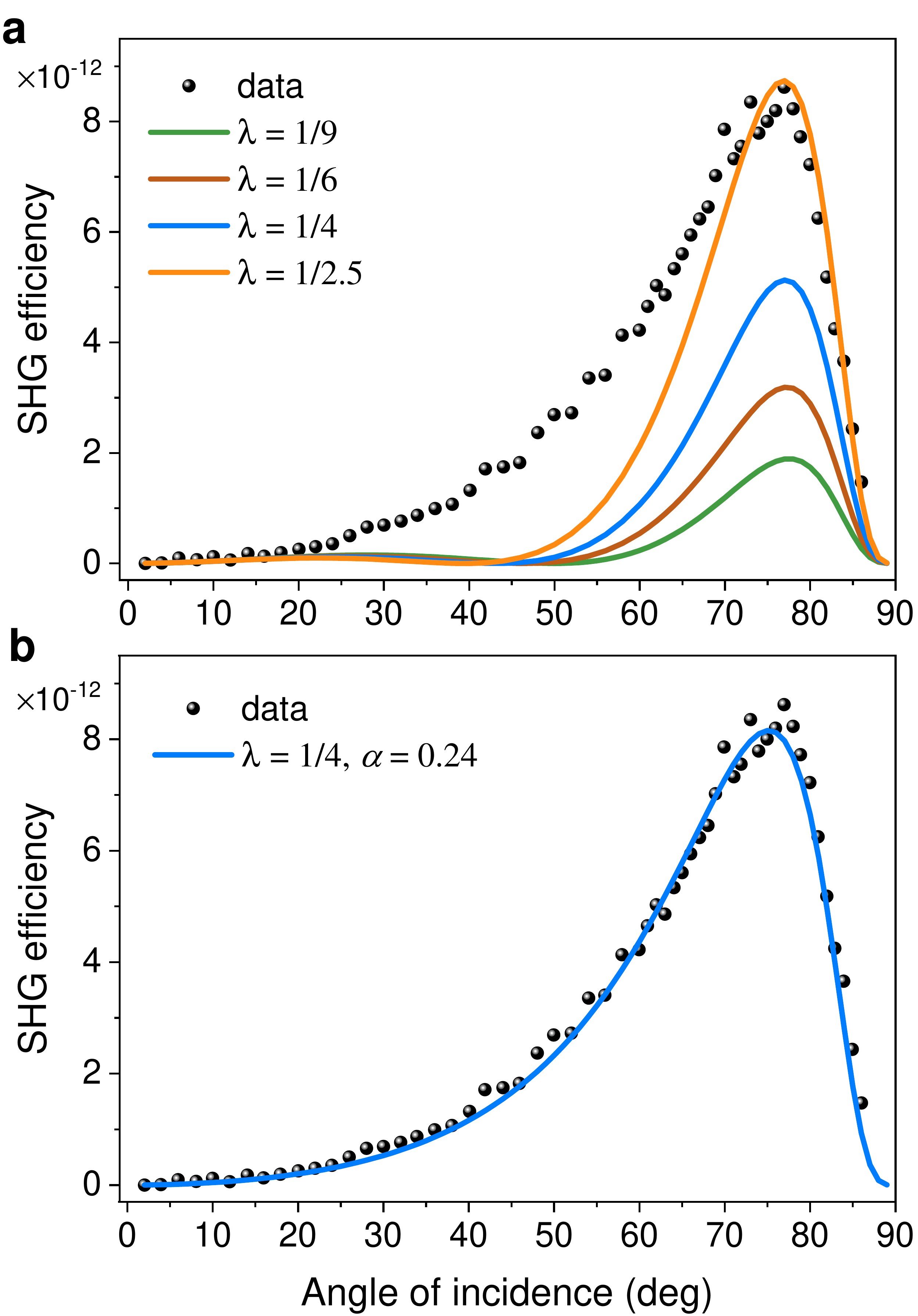}
\caption{\textbf{SHG efficiency as a function of angle of incidence from a 400 nm thick Ag film}. The film is illuminated at $E=1.1653$~eV with a peak intensity $I_0=80$~MW/cm$^2$ and considering the electron damping rate $\hbar \gamma=0.030$~eV; (\textbf{a}) for different values of the spill-out parameter $\lambda$ and assuming $\alpha=1$, the parameter weighting the nonlinear part of the von-Weizs\"acker energy functional. (\textbf{b}) when $\lambda=1/4$ and $\alpha=0.24$. The results are compared against the experimental data taken from Ref. \citenum{ODonnell2005NewJ.Phys.}.}
\label{fig:QHT_ExpData}
\end{figure}

In order to evaluate the relevance of the nonlinear QHT model, we consider a smooth Ag film of 400 nm thickness. We account for core electrons' response with a local permittivity $\varepsilon_\infty = 5.9$ and compare the numerical results with the experimental data already available in the literature \cite{ODonnell2005NewJ.Phys.}. The value of core permittivity $\varepsilon_\infty = 5.9$ is taken from Ref. \citenum{Moaied2017J.Opt.} which is obtained by fitting the classical Drude model with the experimentally measured data. 
The SHG efficiency as a function of angle of incidence for different values of $\lambda$, computed within the full self-consistent nonlinear QHT, is compared against the experimental data, as shown in Fig.~\ref{fig:QHT_ExpData}a.
It can be seen that for smaller $\lambda$ (lower electron spill-out) the efficiency is largely underestimated.
However, as the value of $\lambda$ increases (eventually spill-out increases), the SHG efficiency steps up and for $\lambda=1/2.5$ it coincides well with the peak of the data, although the angular dependence is not thoroughly reproduced.
The experimental data shows a smooth angular dependence while the spectra predicted by the nonlinear QHT exhibit a minimum in the SHG efficiency in the range of $40\degree - 50\degree$.
Although different from the measured data, this trend, and in particular the presence of a minimum, are in agreement with the time-dependent density functional theory calculations \cite{Liebsch1988Phys.Rev.Lett.}.
In order to fit the experimental data, we introduce a phenomenological parameter $\alpha$ weighting the nonlinear part of the von Weizs\"acker energy functional given in equation (\ref{eq:G[n]}) such that:
\begin{equation}
\lambda T_\textrm{W}[n, \nabla n]=\lambda \left[ T_\textrm{W}^\textrm{L}+{\alpha}T_\textrm{W}^\textrm{NL}\right],
\label{eq:alpha}
\end{equation}
where the superscripts $\rm L$ and $\rm NL$ indicate linear and nonlinear contributions, respectively. 
By properly choosing an optimal value of $\alpha$, the experimental data can be nicely approximated qualitatively as well as quantitatively for all angles of incidence.
Figure~\ref{fig:QHT_ExpData}b reports a comparison of the SHG efficiency between the experimental data and the nonlinear QHT when $\lambda=1/4$ and $\alpha=0.24(=1/4.15)$.
It is important to remark that since the full self-consistent QHT does not result in a good agreement with the SHG measurements due to nature of the approximate functionals involved. The introduction of the parameter $\alpha$, however, tunes the nonlinear contributions due to the von Weizs\"acker terms and allows to fit the measured data. As a result, there exists more than one combination of $\lambda$ and $\alpha$ that reproduces experimental data.
The experimental results can also be approximated for other values of $\lambda$ provided that a suitable value of the parameter $\alpha$ is chosen (see Supplementary Note 2). We would also like to point out that the choice of the parameter $\lambda$ and $\alpha$ is not exact and it is a consequence of the non-universality of the kinetic energy functionals, the Thomas-Fermi-von-Weizs\"acker functionals, used to describe the free electron internal energy. Cirac\`i and Della Sala \cite{Ciraci2016Phys.Rev.B} have shown that the spill-out parameter can be fixed by matching the DFT equilibrium density asymptotic decay. The price to pay, however, is giving up self-consistency of the QHT model.

In Ref. \citenum{ODonnell2005NewJ.Phys.}, the authors performed experiments on Ag film for three different pumping-detection polarization conditions: (i) TM-incident and TM-detected, (ii) TE-incident and TM-detected, and (iii) $45\degree$ TE/TM-incident and TE-detected. The results shown in Fig. \ref{fig:QHT_ExpData} correspond to the first case, i.e., when the incident and detected signals are TM-polarized. Note that in this condition, most of the surface contributions are activated. The SHG efficiency spectra for the other polarizations are presented in the Supplementary Note 3.  
\subsection*{SHG resonances in Ag film with a dielectric coating}\label{subsec:OL_AgFilm}
\noindent We have seen in the previous section that the Na films exhibit a strong enhancement in the SHG spectra due to the presence of resonances induced by the electron density decays at the metal surface.
Contrarily to the Na case, however, we could not see any resonance structures neither in the linear nor in the SHG spectra for Ag films.
The lack of such resonances is due to the combination of two factors.
On the one hand, a larger ion density in Ag (smaller Wigner-Seitz radius) compared to Na, leads to a greater work function that keeps the electron tighter at the metal surface, producing a much smaller spill-out.
As we have seen by varying $\lambda$, a faster decay of the electron density pushes the resonance to higher energies.
On the other hand, the presence of a polarizable background in Ag (due to interband transitions) shifts the effective plasma frequency to lower energies. 
The result is that the Ag film becomes \textit{transparent} before any multipolar resonance can be excited, completely suppressing the behavior observed for Na.

\begin{figure*}
\centering
\includegraphics[width=0.95\linewidth]{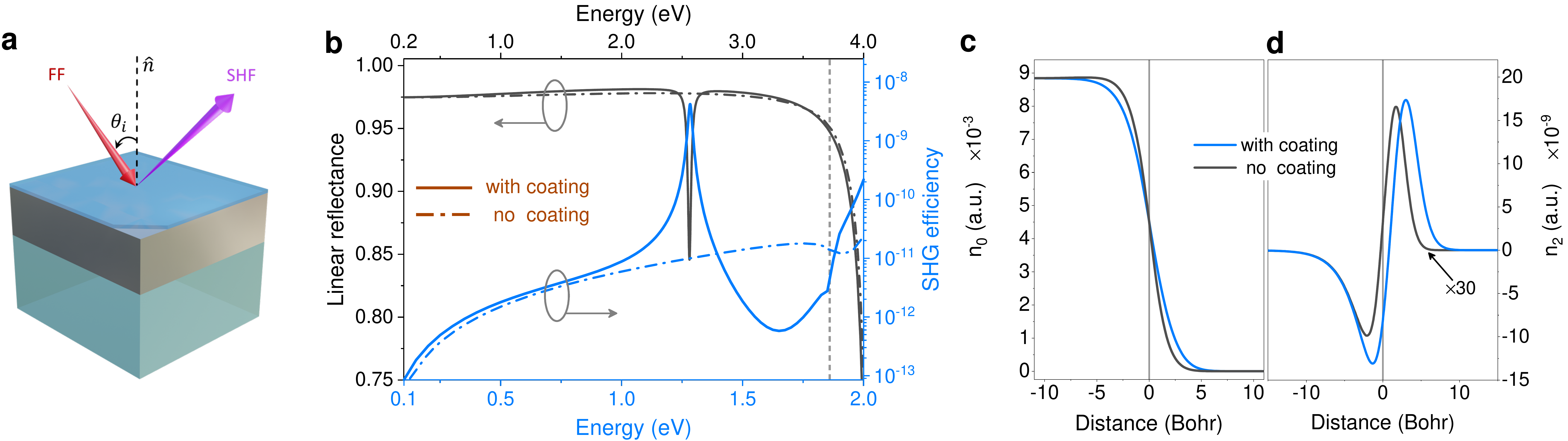}
\caption{\textbf{Linear and nonlinear response of a Ag film coated with a sub-wavelength thin layer of a dielectric material}. (\textbf{a}) Simulation setup for SHG from the Ag film placed on a glass substrate. (\textbf{b}) Linear reflectance and SHG efficiency from a 400 nm thick Ag film coated with a 10~nm thick dielectric layer with dielectric constant $\varepsilon_r=5.9$ (solid lines). The film is excited with a transverse magnetic polarized plane wave impinging at an angle $\theta_i=75\degree$ by considering the spill-out parameter $\lambda=1/4$ and the weight factor $\alpha=0.24$. The results for the Ag film without any coating are also shown for comparison (broken curves). The vertical dotted grey line represents the bulk plasma frequency. (\textbf{c}) Self-consistent equilibrium density $n_0$ and (\textbf{d}) induced charge density $n_2$ at $E=1.28$ eV for coated and uncoated films, plotted along the normal direction to the metal interface. The vertical solid grey line marks the metal interface.}
\label{fig:OLAg}
\end{figure*}

One possible way to circumvent this issue is to increase the electron spill-out from the Ag surface by reducing the work function at the metal interface.
The work function predicted for the QHT can be expressed as\cite{Yan2015Phys.Rev.B}:
\begin{equation}
    W=\lambda \frac{\hbar k_D^2}{2 m_e}
\end{equation}
where $k_D$ describes the decay of the charge density proportional to $\sim e^{-k_Dx}$.
It can be seen from the above equation that when $k_D$ is smaller, i.e., the decay is slower, which implies that the spill-out is higher and therefore the work function decreases.
The potential barrier at the metal surfaces can be lowered by reducing the difference in the dielectric constant between the metal background and the surround region.
This can be obtained by modifying the dielectric environment surrounding the metal.
In order to do so, we coat the film with a sub-wavelength thin layer of a dielectric material with a dielectric constant $\varepsilon_r$, as depicted in Fig.~\ref{fig:OLAg}a.
The impact of the dielectric coating material on the equilibrium charge density is shown in the Supplementary Fig.~3 (see Supplementary Note~4).
The linear reflectance and SHG efficiency spectra for a 400 nm Ag film, coated with a 10 nm thick layer of a dielectric material with $\varepsilon_r=\varepsilon_\infty$ and placed on a glass substrate with refractive index $n=1.45$, are plotted in Fig.~\ref{fig:OLAg}b.
Note that 
the figure also shows, for the sake of comparison, the spectra for the bare Ag film without any coating (as in Fig.~\ref{fig:Slab_Geometery}).
We observe that in the case of dielectric coating of Ag film, the linear response shows a dip in the reflectance which will result into a very strong resonance in the SHG efficiency, however, the Ag film without a coating does not show any resonance structure.
This result is, in fact, very interesting as the spill-out induced resonances, which are suppressed in the commonly used setups in plasmonics, could be excited through a dielectric coating and, therefore, can be exploited for enhancing SHG.

The reason why the dielectric-coated Ag surface supports such resonant feature can be better understood by looking at the charge densities of the coated and uncoated Ag films. Figure~\ref{fig:OLAg}c presents a comparison between the equilibrium charge density of a clean and dielectric-coated Ag film. It can be observed that that due to presence of the dielectric coating, the electron spill-out is more pronounced as compared to the uncoated slab.
The dielectric material actually provides a higher electron density of states channel, allowing for the electrons to move further away from the metal surface, and as a consequence of this higher spill-out, the resonances can be observed in the linear and nonlinear spectra. Figure~\ref{fig:OLAg}c presents the electron density both for the coated and uncoated films at $\lambda=1/4$. 
Thus, the dielectric coating significantly assists the electron to spill-out of the metal surface and consequently producing a strong enhancement in the SHG efficiency.
The induced charge density at the SHG, plotted at $E=1.28$ eV, for the coated and uncoated case is given in Fig.~\ref{fig:OLAg}d.
In both cases, the centroid of the charge density lies outside of the positive background edge. Although the spatial variation is quite similar, the induced charge density for the coated film is comparatively larger in width and magnitude, and is shifted farther away from the metal surface.

\begin{figure*}
\centering
\includegraphics[width=0.8\linewidth]{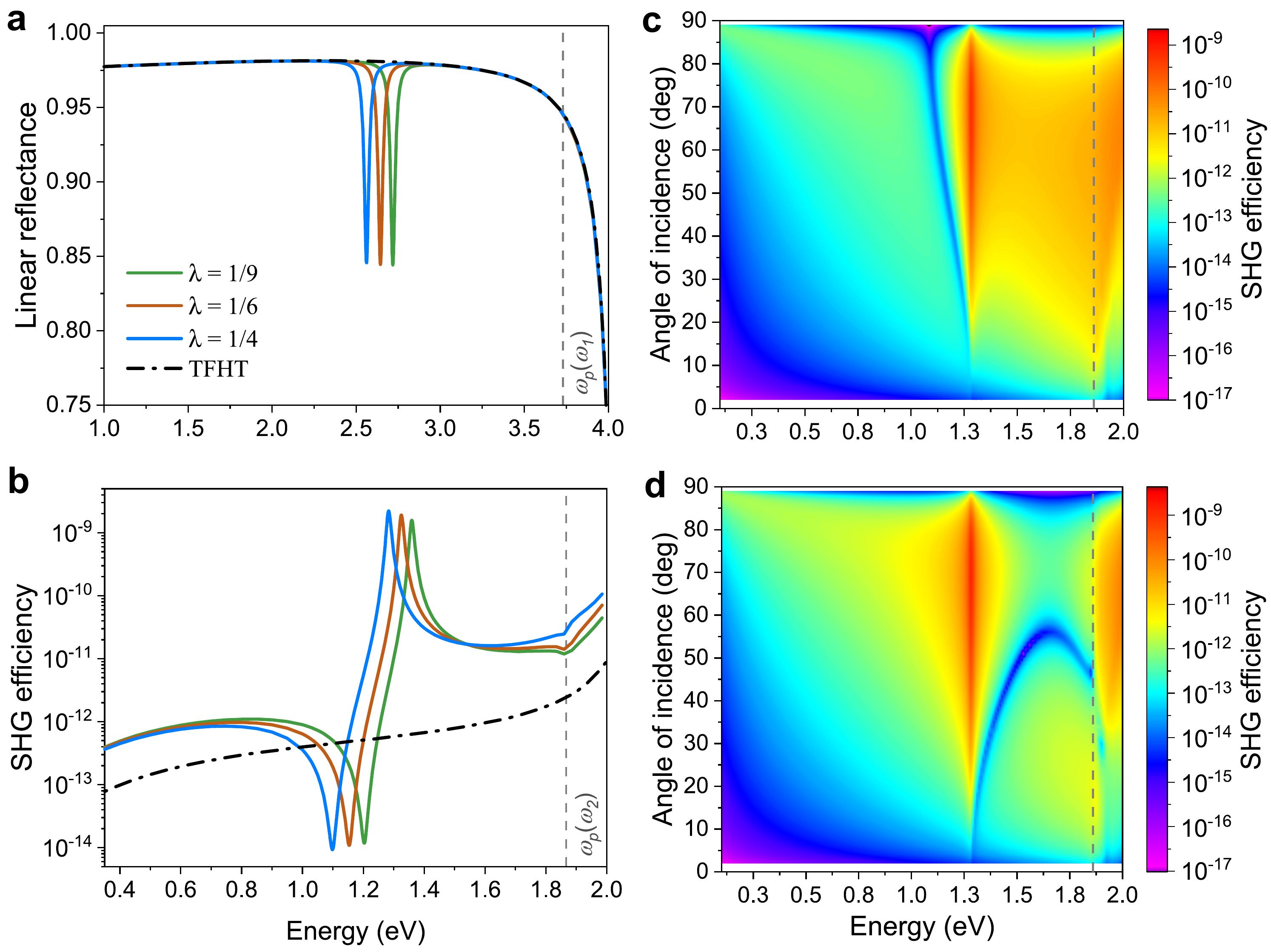} 
\caption{\textbf{Optical reflection from the dielectric-coated Ag film considering different amount of electron spill-out from the metal surface}. (\textbf{a}) Linear reflectance and (\textbf{b}) SHG efficiency as a function of exciting energy (in eV) of the incident field for different values of the spill-out parameter $\lambda$ and considering the weight factor $\alpha=1$. The film is coated with a 10~nm thick dielectric layer with a dielectric constant $\varepsilon_r=5.9$ and excited with a transverse magnetic polarized light incident at an angle $\theta_i=75\degree$. The results are compared against the Thomas-Fermi hydrodynamic theory (TFHT) spectra represented by the dash-dotted black curves. The vertical grey lines indicate the bulk plasma frequency $\omega_p$ at the fundamental and generated harmonic. SHG efficiency of the dielectric-coated Ag slab, for $\lambda=1/4$ considering (\textbf{c}) $\alpha=1$ and (\textbf{d}) $\alpha=0.24$, plotted as a function of exciting energy and angle of incidence.}
\label{fig:MapPlots}
\end{figure*}

It is interesting to note that a similar behavior can be observed for different amount of the electron spill-out and it is not the product of a specific choice of parameters. 
In Figs.~\ref{fig:MapPlots}a and ~\ref{fig:MapPlots}b, we report linear and SHG spectra, respectively, for the dielectric-coated Ag slab for different values of $\lambda$ and compare results against the TFHT (with no spill-out).
For each $\lambda$, we observe a large resonance in the SHG spectrum which is related to the dip in the corresponding linear reflectance.
Neglecting the electron spill-out from the metal surface, as in the TFHT, prevents such features from appearing neither in SHG nor in linear spectra.    
The enhancement in the SHG efficiency is therefore strongly depends on the surface conditions. About three decades ago, monochromatic-light experiments were performed on dielectric ($\textrm{C}_{60}-$ and $\textrm{SiO}_x-$) coated Ag island films to study the SHG and the sensitivity of the generated harmonic to the surface properties was emphasized. An order of magnitude enhancement in the SH response was observed in the coated Ag island films as compared to the uncoated ones \cite{Aktsipetrov1993_PLA}.

The SHG spectra of Fig.~\ref{fig:MapPlots}b display a noticeable minimum just before the peak resonance for each $\lambda$, which is not seen in the situation shown in Fig.~\ref{fig:OLAg}b.
To examine this feature, we consider $\lambda=1/4$ and compute the SHG efficiency as a function of the angle of incidence and the fundamental energy for different values of $\alpha$.
We find that when $\alpha=1$, there exists a minimum at almost all angles of incidence (Fig.~\ref{fig:MapPlots}c) whereas for $\alpha=0.24$, there is no such minimum before the resonance, nevertheless, in this scenario a minimum also appears but after the resonance (Fig.~\ref{fig:MapPlots}d).
In the latter case, the minimum is not present for all angles of incidence and since the spectrum shown in Fig.~\ref{fig:OLAg}b is plotted at $\theta_i=75\degree$, therefore, no such minimum is detected there.
This minimum in the SHG efficiency can be associated to the interplay between the different nonlinear contributions that involve the first-order and second-order terms of the von Weizs\"acker energy functional, respectively. While varying $\lambda$ has a simultaneous effect on both terms, varying $\alpha$ modifies their respective weights, modifying the condition for which the two contributions cancel out.  

\begin{figure}
\centering
\includegraphics[width=0.85\linewidth]{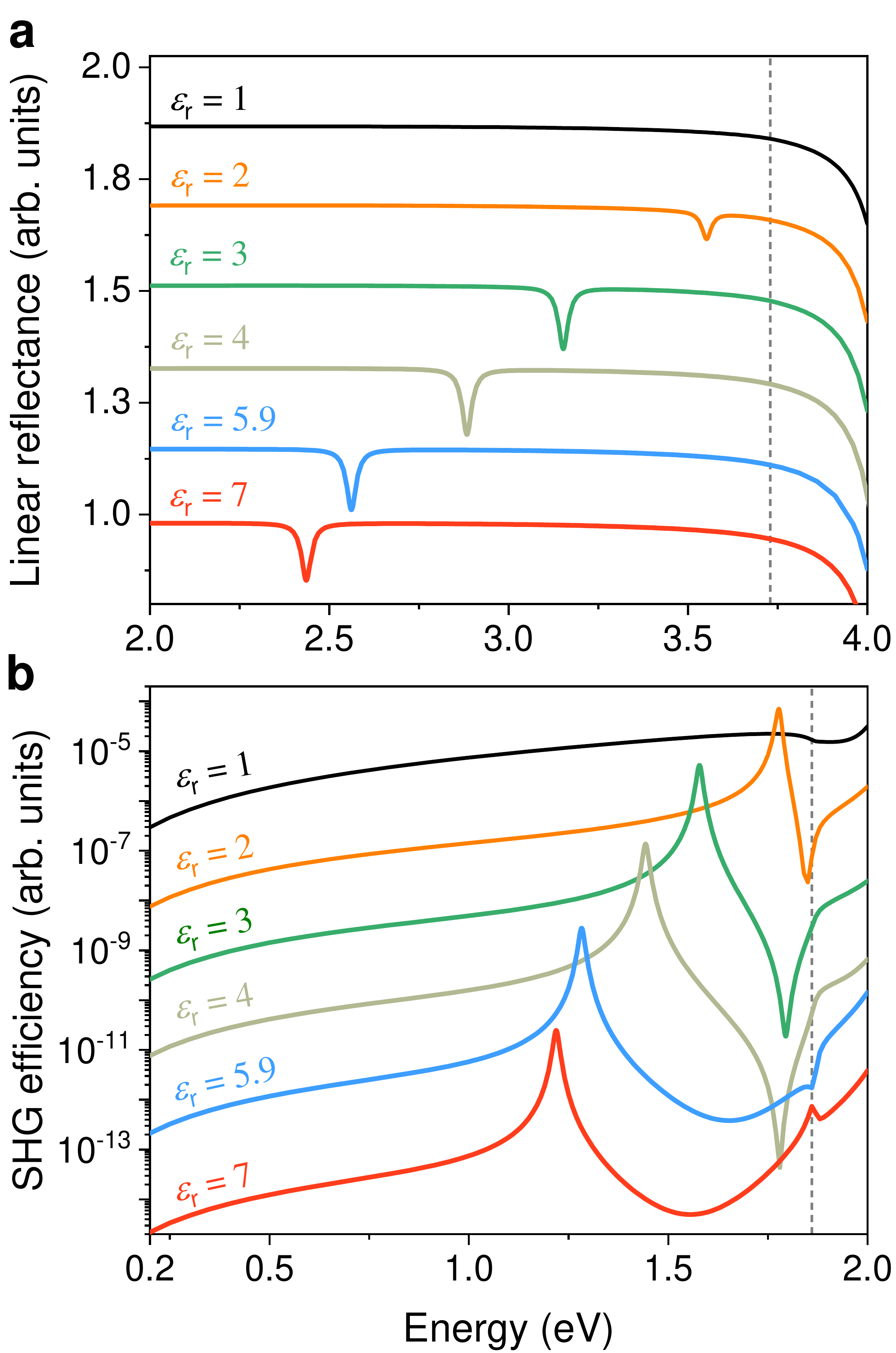}
\caption{\textbf{Impact of the dielectric constant of the coating material on the reflection properties of Ag film}. (\textbf{a}) Linear reflectance and (\textbf{b}) SHG efficiency as a function of exciting energy plotted at an incident angle $\theta_i=75\degree$ for various values of the dielectric constant $\varepsilon_r$ considered for the coating of Ag film. The result for the clean Ag film without any coating $(\varepsilon_r=1)$ is also reported. The linear and nonlinear spectra are arbitrarily shifted in the vertical direction for the sake of a better visualization.}
\label{fig:epsrOL}
\end{figure}

So far, we have fixed the refractive index of the coating material such that it corresponds to $\varepsilon_\infty$ in order to reproduce a similar condition to Na where $\varepsilon_\infty=1$.
It would be interesting at this point, studying the influence of the dielectric constant of the coating material on the SHG efficiency.
In Fig.~\ref{fig:epsrOL}, we report numerical results for linear response and SHG of the Ag film considering $\lambda=1/4, \alpha=0.24$ and various values of the dielectric constants $\varepsilon_r$ of the coating material.
The spectra for silver-air interface ($\varepsilon_r=1$, i.e., without any coating) are also shown for completeness.
Contrarily to the Ag-air interface, when $\varepsilon_r \neq 1$, a dip in the linear reflectance can be seen for each $\varepsilon_r$, although it becomes less intense for the smaller values of $\varepsilon_r$ and completely disappears for $\varepsilon_r=1$, as shown in Fig.~\ref{fig:epsrOL}a.
This structure in the linear spectrum yields a large efficiency of the generated signal, as depicted in Fig.~\ref{fig:epsrOL}b.
It is interesting to note that as the value of the dielectric constant of the coating layer increases, the resonance shifts towards the lower energies.
This result, in fact, could be very important in many optical applications as the resonance of the SHG process can be tuned by coating the Ag film with a suitable dielectric material.
A minimum in the SHG spectra for a lower value of the dielectric constants (with the exception of $\varepsilon_r=1$) is also visible in the SHG spectra, which appears after the resonance in the vicinity of bulk plasma frequency and corresponds to the similar situation as was discussed in Fig.~\ref{fig:MapPlots}b. However, this minimum vanishes for the higher values of the dielectric constant. %
\section*{Discussion}
\noindent We have presented a theoretical microscopic model to probe second-order nonlinearities at metal surfaces.
Our approach is based on the QHT, which can efficiently handle the realistic profiles of the ground-state  electron density.   
The nonlinear QHT predicts a very strong resonance in the SHG spectra that can boost the conversion efficiency by several orders of magnitude.
This resonance shows a strong dependence on the spill-out of the electron density from the metal surface and is related to the excitation of multipolar surface plasmons in the linear response.
Interestingly, these spill-out induced resonances can be tuned by controlling decay of the electron density from the metal surface with the aid of a dielectric coating.

Our results suggest that an accurate description of the spatial variation of charge density at a metal surface is crucial in characterizing an accurate nonlinear optical response.
Although we have limited our study to films, the spill-out induced enhancement of SHG efficiency could be combined with plasmon-based field enhancement techniques to further increase nonlinear conversion rates. 
Moreover, because the nonlinear process is confined to the metal surface, such processes could be embedded in a hybrid metal-dielectric waveguide configuration that allows the build up of the nonlinear signals. 
We believe that the analysis presented in this article will be a source of motivation for further analytical and experimental research.   
\section*{Methods}\label{sec:QHT}
\subsection*{Nonlinear quantum hydrodynamic theory}
\noindent Materials with a very high density of free charges can be treated as charged fluids (in a positively charged background) and their optical properties can be characterized under the hydrodynamic description.
The full nonlinear hydrodynamic equation of motion for an electronic system under the influence of electromagnetic fields $\boldsymbol{\mathrm{E}}$ and $\boldsymbol{\mathrm{B}}$ can be expressed as \cite{Ciraci2016Phys.Rev.B}:
\begin{equation}
 m_e\left(\frac{\partial }{\partial t}+\boldsymbol{\mathrm{v}}\cdot \mathrm{\nabla }+\gamma \right)\boldsymbol{\mathrm{v}}=-e\left(\boldsymbol{\mathrm{E}}+\boldsymbol{\mathrm{v}}\times \boldsymbol{\mathrm{B}}\right)-\mathrm{\nabla }\frac{\delta G[n]}{\delta n}, 
\label{eq:QHTgeneral}
\end{equation}
where $m_e$ is the electron mass, $\boldsymbol{\mathrm{v}}$ is the electron velocity field, $\gamma$ is the damping rate and $e$ is the absolute value of the electron charge; $n({\bf r})$ represents the electron density and $G[n]= T_{\textrm{TF}}[n]+\lambda T_{\textrm{W}}[n, \nabla n]+E_{\textrm{xc}}[n]$ is the energy functional incorporating the TF kinetic energy, the gradient-dependent correction term and the exchange-correlation energy functional respectively.
In specifying the energy functionals in equation (\ref{eq:G[n]}), we do not include the effect of nonlocal broadening (see Ref. \citenum{Ciraci2017Phys.Rev.B} for more details) to avoid further numerical complexity.
Considering the electron current $\boldsymbol{\mathrm{J}}=-ne\boldsymbol{\mathrm{v}}$, equation (\ref{eq:QHTgeneral}) can be expressed as:
\begin{equation}
\begin{split}
\frac{\partial \boldsymbol{\mathrm{J}}}{\partial t}+\gamma \boldsymbol{\mathrm{J}}=\frac{e^2n}{m_e}\boldsymbol{\mathrm{E}}&-\frac{e}{m_e}\boldsymbol{\mathrm{J}}\times \boldsymbol{\mathrm{B}}+\frac{en}{m_e}\mathrm{\nabla }\frac{\delta G[n]}{\delta n}+ \\ 
&+\frac{1}{e}\left(\frac{\boldsymbol{\mathrm{J}}}{n}\mathrm{\nabla}\cdot \boldsymbol{\mathrm{J}}- \boldsymbol{\mathrm{J}}\cdot \mathrm{\nabla}\frac{\boldsymbol{\mathrm{J}}}{n}\right).
\end{split}
\label{eq:standard_J}
\end{equation}
A similar equation can also be directly derived from the single-particle Kohn-Sham equation \cite{Ciraci2017Phys.Rev.B}.

We proceed by writing the fields as the sum of few harmonics:
\begin{eqnarray}
n(\boldsymbol{\mathrm{r}},t)&=&n_0(\boldsymbol{\mathrm{r}})+n_1(\boldsymbol{\mathrm{r}})e^{-i\omega_1 t}+n_2(\boldsymbol{\mathrm{r}})e^{-i\omega_2 t}+ \textrm{c.c.}, \label{eq:n_expand}\\
\boldsymbol{\mathrm{E}}(\boldsymbol{\mathrm{r}},t)&=&\boldsymbol{\mathrm{E}}_0(\boldsymbol{\mathrm{r}})+\boldsymbol{\mathrm{E}}_1(\boldsymbol{\mathrm{r}})e^{-i\omega_1 t}+\boldsymbol{\mathrm{E}}_2(\boldsymbol{\mathrm{r}})e^{-i\omega_2 t} + \textrm{c.c.}, \label{eq:E_expand}\\
\boldsymbol{\mathrm{J}}(\boldsymbol{\mathrm{r}},t)&=&\boldsymbol{\mathrm{J}}_1(\boldsymbol{\mathrm{r}})e^{-i\omega_1 t}+\boldsymbol{\mathrm{J}}_2(\boldsymbol{\mathrm{r}})e^{-i\omega_2 t} +\textrm{c.c.}, \label{eq:J_expand}\\
\boldsymbol{\mathrm{B}}(\boldsymbol{\mathrm{r}},t)&=&\boldsymbol{\mathrm{B}}_1(\boldsymbol{\mathrm{r}})e^{-i\omega_1 t}+\boldsymbol{\mathrm{B}}_2(\boldsymbol{\mathrm{r}})e^{-i\omega_2 t}+\textrm{c.c.}, \label{eq:B_expand}
\end{eqnarray}
where $\omega_2=2\omega_1$ and $\textrm{E}_0$ and $n_0$ represent the steady-state electric field and ground-state electron density. The subscripts $1$ and $2$ indicate the field quantities at the fundamental and the second harmonic, respectively. Assuming the undepleted pump approximation, that is, the fundamental field is not affected by the conversion process, and recalling that $\boldsymbol{\mathrm{J}}=\dot{\boldsymbol{\mathrm{P}}}$, in view of equations (\ref{eq:n_expand})-(\ref{eq:B_expand}), equation (\ref{eq:standard_J}) results into following set of equations:
%
\begin{align}
&-\frac{en_0}{m_e} \nabla \left(\frac{\delta G[n]}{\delta n} \right)_1-\left(\omega_1^2+i \gamma\omega_1 \right) \boldsymbol{\mathrm{P}}_1=\varepsilon_0 \omega_p^2 \boldsymbol{\mathrm{E}}_1, \label{eq:FF}\\
&-\frac{en_0}{m_e} \nabla \left(\frac{\delta G[n]}{\delta n} \right)_2-\left(\omega_2^2+i \gamma\omega_2 \right) \boldsymbol{\mathrm{P}}_2=\varepsilon_0 \omega_p^2\boldsymbol{\mathrm{E}}_2+\boldsymbol{\mathrm{S}}_{\mathrm{NL}}, \label{eq:SH}
\end{align}
where $\omega_p(\boldsymbol{\mathrm{r}})=\sqrt{e^2 n_0(\boldsymbol{\mathrm{r}})/(m_e \varepsilon_0)}$ is the space-dependent plasma frequency and the nonlinear source term is:
\begin{equation}
\begin{split}
\boldsymbol{\mathrm{S}}_{\mathrm{NL}}=&\frac{e^2n_1}{m_e}  \boldsymbol{\mathrm{E}}_1 +i\omega_1 \frac{\mu_0 e}{m_e}\boldsymbol{\mathrm{P}}_1\times \boldsymbol{\mathrm{H}}_1 +\\
& - \frac{{\omega _1^2}}{{e{n_0}}}\left( {{{\bf{P}}_1}\nabla  \cdot {{\bf{P}}_1} + {{\bf{P}}_1} \cdot \nabla {\bf{P}}_1 - \frac{{\bf{P}}_1\cdot\nabla n_0}{n_0}{\bf{P}}_1} \right)+ \\
&+ \frac{en_1}{m_e}\nabla \left(\frac{\delta G[n]}{\delta n} \right)_1+ \frac{en_0}{m_e}\nabla \left(\frac{\delta G[n]}{\delta n} \right)^{\mathrm{NL}}_2.
\end{split}
\label{eq:S_NL}
\end{equation}
Mathematical expressions for the linear potentials $\left(\frac{\delta G[n]}{\delta n} \right)_{1,2}$ are reported in detail in Ref. \citenum{Ciraci2016Phys.Rev.B} whereas expressions for the nonlinear potentials are derived in the Supplementary Note 1. 
In addition to the classical nonlinear hydrodynamic terms \cite{Scalora2010Phys.Rev.A, Ciraci2012Phys.Rev.B}, i.e., Coulomb, Lorentz and convective terms (note that there is one extra convective term, $\propto\nabla n_0/n_0$, arising from the spatial dependence of the electron density), two  quantum terms emerge as discussed in the Results section.
\subsection*{Self-consistent equilibrium charge density}
In implementing our nonlinear QHT, we follow the jellium approximation \cite{Brack1993Rev.Mod.Phys.}, which assumes that the electrons in a metal are confined by a constant positive background charge $n^+=(\frac{4}{3}\pi r_s^3)^{-1}$, where $r_s$ indicates the Wigner-Seitz radius ($r_s=4$ for Na and $r_s=3$ for Ag).
The space-dependent equilibrium charge density can be calculated by combining the zero-\textit{th} order QHT equation with the Gauss's law, which leads to the following nonlinear static equation \cite{Ciraci2017Phys.Rev.B}:
\begin{equation}
\nabla \varepsilon_\infty(\boldsymbol{\rm r}) \cdot \nabla \left(\frac{\delta G}{\delta n} \right)_{n=n_0}+\frac{e^2}{\varepsilon_0}\left(n_0-n^+ \right)=0,
\label{eq:scn0}
\end{equation}
where $\varepsilon_0$ symbolizes permittivity of free space. $\varepsilon_\infty$ is the core dielectric constant presenting a local contribution to the metal permittivity.

We solve equations (\ref{eq:FF})--(\ref{eq:scn0}) coupled to two frequency-dependent wave equations in a finite-element-method based numerical simulator \textsc{comsol} Multiphysics \cite{comsol}, which offers a quite flexible implementation of these expressions. The numerical implementation of these equations is given in the Supplementary Note 5.
\section*{Data Availability}
The data that support the findings of this study are available from the corresponding author upon reasonable request.
\section*{Code Availability}
The codes that support the findings of this study are available from the corresponding author upon reasonable request.
\def\bibsection{\section*{References}}
\bibliography{myLibrary}
\begin{acknowledgements}
\noindent C. C. thanks Prof. John Sipe for valuable insights and discussions.
The authors thank the Air Force Office of the Scientific Research for supporting this work under the award number FA9550-17-1-0177.
\end{acknowledgements}
\section*{Author Contributions}
\noindent C.C. developed the theory; M.K. set up the numerical model and performed the simulations; C.C. and M.K. analyzed the numerical data and wrote the manuscript; C.C. supervised the project.
\section*{Competing Interests}
The authors declare no competing interests.
\section*{Additional information}
\textbf{Supplementary information} is available for this paper at xxx.xxx.xxx
\vfill
\newpage
\begingroup
\renewcommand{\color}[1]{}
%
\eject
\end{document}


\title{\large \centering  Supplementary Material: Enhancing second-harmonic generation with electron spill-out at metallic surfaces}

\author{Muhammad Khalid}
\email{muhammad.khalid@iit.it}

\author{Cristian Cirac\`i}
\email{cristian.ciraci@iit.it}
\date{\today}
\affiliation{Istituto Italiano di Tecnologia, Center for Biomolecular Nanotechnologies, Via Barsanti 14, 73010 Arnesano, Italy.}

\maketitle

\thispagestyle{plain} 

\setcounter{equation}{0}
\setcounter{figure}{0}
\setcounter{table}{0}
\setcounter{page}{1}
\makeatletter

\setcounter{section}{0}
\renewcommand{\thesection}{S\arabic{section}}

\renewcommand{\theequation}{S\arabic{equation}}
\renewcommand{\figurename}{Supplementary Figure}


\begin{flushleft}
\section*{Supplementary Note 1: Nonlinear terms of the energy functionals}\label{supp:NL}
\end{flushleft}
\noindent Recalling the energy functional $G[n]$, representing the total internal energy of an electronic system (see equation~(1) in the main text), 
\begin{equation}
G[n] = T_{\textrm{TF}}[n]+\lambda T_{\textrm{W}}[n, \nabla n]+E_{\textrm{xc}}[n],
\label{eq:G[n]}
\end{equation}
Here we report the derivation of the nonlinear terms of the energy functionals appearing in the above equation. 
The functional derivative of the terms in Supplementary Equation~(\ref{eq:G[n]}) can be expressed as \cite{Ciraci2016Phys.Rev.B},
\begin{eqnarray}
\frac{\delta T_{\rm TF}}{\delta n}&=&\left(E_h a_0^2 \right) \frac{5}{3} c_{\rm TF} n^{2/3}, \label{eq:dTF}\\
\frac{\delta T_{\rm W}}{\delta n}&=&\left(E_h a_0^2 \right) \frac{1}{8} \left(\frac{\nabla n \cdot \nabla n}{n^2}-2\frac{\nabla^2 n}{n} \right), \label{eq:dvW}\\
\frac{\delta E_{\rm xc}}{\delta n}&=&\varepsilon_{\rm xc}+n\frac{d\varepsilon_{\rm xc}}{dn}, \label{eq:dXC}
\end{eqnarray}
where,
\begin{eqnarray}
\varepsilon_{\rm xc}&=&-c_{\rm x} n^{1/3}+\varepsilon_{\rm c}, \\
\varepsilon_{\rm c}&=&
\begin{cases}
a_c \ln r_s + b_c+ c_c r_s \ln r_s+d_c r_s, & r_s<1, \\
\frac{\gamma_c}{1+\beta_1 r_s^{1/2}+\beta_2 r_s}, & r_s\ge1, \label{eq:e_c}
\end{cases}
\end{eqnarray}
and $E_h=\frac{\hbar^2}{m_e a_0^2}$ is the Hartree energy, $a_0$ represents the Bohr radius, $r_s$ is the Wigner-Seitz radius, $c_{\rm TF}=\frac{3}{10}\left(3\pi^2\right)^{1/3}$ and $c_{\rm x}=\frac{3}{4}\left(\frac{3}{\pi} \right)^{1/3}$. The coefficients appearing in Supplementary Equation~(\ref{eq:e_c}), as given in Supplementary Ref.~\citenum{Ciraci2017Phys.Rev.B}, are: $a_c=0.0311, b_c=-0.048, c_c=0.002, d_c=-0.0116, \gamma_c=-0.1423, \beta_1=1.0529$ and $\beta_2=0.3334$. Electron density as the sum of few harmonics can be expressed as:
\begin{equation}
n=n_0 + \sum_j n_j e^{-i\omega_jt}+{\rm c.c.}, \qquad \text{where,}\quad j=1,2 
\end{equation}
For mathematical convenience, dropping the summation operator and writing the above equation simply as (following the Einstein summation notation):
\begin{equation}
n=n_0 + n_je^{-i\omega_jt}+{\rm c.c.} \label{eq:n_j}
\end{equation}
Second-order contribution of $\frac{\delta T_{\rm TF}}{\delta n}$ can be obtaind by substituting Supplementary Equation~(\ref{eq:n_j}) into Supplementary Equation~(\ref{eq:dTF}) and applying the Binomial expansion, we get,
\begin{equation}
\frac{\delta T_{\rm TF}}{\delta n}=\left(E_h a_0^2 \right) \frac{5}{3} c_{\rm TF} \left[n_0^{2/3}+\frac{2}{3}n_0^{-1/3}\left(n_je^{-i\omega_jt}\right)-\frac{1}{9}n_0^{-4/3}\left(n_je^{-i\omega_jt}\right)^2 +{\rm c.c.}+\dots \right],
\end{equation}
Focusing only on the second-order terms and neglecting the higher harmonics ($\omega>\omega_2$), we are left with:
\begin{equation}
\left(\frac{\delta T_{\rm TF}}{\delta n} \right)_2^{\textrm{NL}}=-\frac{5}{27}c_{\rm TF}n_0^{-3/4}n_1^2. 
\end{equation}
Now, the nonlinear terms of the $T_{\rm W}$ can be derived by inserting Supplementary Equation~(\ref{eq:n_j}) into Supplementary Equation~(\ref{eq:dvW}), as:
\begin{equation}
\frac{\delta T_{\rm W}}{\delta n}=\left(E_h a_0^2 \right) \frac{1}{8} \left[\frac{\nabla \left(n_0 + n_je^{-i\omega_jt} \right) \cdot \nabla \left(n_0 + n_je^{-i\omega_jt} \right)}{\left(n_0 + n_je^{-i\omega_jt} \right)^2}-2\frac{\nabla^2 \left(n_0 + n_je^{-i\omega_jt} \right)}{n_0 + n_je^{-i\omega_jt}} \right].
\end{equation}
Expanding $\left(n_0 + n_je^{-i\omega_jt}\right)^{-2}$ in the above equation and doing some algebra results in the following nonlinear contribution:
\begin{equation}
\left(\frac{\delta T_{\textrm{W}}}{\delta n} \right)^{\textrm{NL}}_2 =\frac{\nabla n_1 \cdot \nabla n_1 }{n_0^2}-4\frac{\nabla n_0 \cdot \nabla n_1}{n_0^3}n_1+2\frac{\nabla^2 n_1}{n_0^2}n_1+3\frac{\left|\nabla n_0 \right|^2}{n_0^4} n_1^2-2\frac{\nabla^2 n_0}{n_0^3}n^2_1.
\end{equation} 
Similarly, the nonlinear terms for exchange-correlation potential can be obtained by using Supplementary Equation~(\ref{eq:n_j}) into Supplementary Equation~(\ref{eq:dXC}), resulting in the following expression:
\begin{equation}
\left(\frac{\delta E_{\mathrm{xc}}}{\delta n} \right)^{\textrm{NL}}_2 = \frac{1}{2} \left(3 \frac{d^2 \mathrm{\epsilon_x}}{dn^2}+n \frac{d^3 \mathrm{\epsilon_x}}{dn^3}+3 \frac{d^2 \mathrm{\epsilon_c}}{dn^2}+n \frac{d^3 \mathrm{\epsilon_c}}{dn^3} \right)_{n=n_0} n_1^2,
\end{equation}
where,
\begin{eqnarray*}
\frac{d^2 \mathrm{\epsilon_x}}{dn^2} &=& \frac{2}{9} c_\mathrm{x} n^{-5/3}, \\
\frac{d^3 \mathrm{\epsilon_x}}{dn^3} &=& \frac{10}{27} c_\mathrm{x} n^{-8/3}, \\
\frac{d^2 \mathrm{\epsilon_c}}{dn^2} &=&
\begin{cases}
&\frac{1}{3} \left[\frac{r_s}{3} \left(5c_c+4d_c+4c_c \mathrm{ln} r_s \right)+a_c \right]n^{-2},  \quad r_s<1, \\
& \frac{\gamma_c}{9}\left[\frac{\frac{7\beta_1}{4}r_s^{1/2}+4\beta_2 r_s}{\left(1+\beta_1 r_s^{1/2}+\beta_2 r_s \right)^2}+\frac{\left(\frac{\beta_1}{2}r_s^{1/2}+\beta_2 r_s\right)^2}{\left(1+\beta_1 r_s^{1/2}+\beta_2 r_s \right)^4} \right]n^{-2}, \quad r_s \geq 1,
\end{cases} \\
\frac{d^3 \mathrm{\epsilon_c}}{dn^3} &=&
\begin{cases}
&-\frac{1}{9} \left[\left(13c_c+\frac{28}{3}d_c+\frac{28}{3}c_c \mathrm{ln} r_s \right)r_s+6a_c \right]n^{-3}, \quad r_s < 1, \\
&\frac{\gamma_c}{9}\left[\frac{\frac{7\beta_1}{2}r_s^{1/2}+8\beta_2 r_s}{\left(1+\beta_1 r_s^{1/2}+\beta_2 r_s \right)^2}+\frac{\frac{7}{24}\beta_1 r_s^{1/2}+\frac{4}{3}\beta_2 r_s-\frac{7}{24}\beta_1 r_s -\frac{7}{8}\beta_1\beta_2 r_s^{2/3}-\frac{4}{3}\beta_2^2 r_s^2}{\left(1+\beta_1 r_s^{1/2}+\beta_2 r_s \right)^4} \right. + \\
&+\left. \frac{2 \left(\frac{\beta_1}{2}r_s^{1/2}+\beta_2 r_s\right)^2}{\left(1+\beta_1 r_s^{1/2}+\beta_2 r_s \right)^4}-\frac{\left(\frac{\beta_1}{2}r_s^{1/2}+\beta_2 r_s\right) \left(-\frac{1}{6}\beta_1 r_s -\frac{2}{3}\beta_2 r_s +\frac{1}{6}\beta_1^2 r_s +\frac{1}{2}\beta_1\beta_2 r_s^{3/2}+\frac{2}{3}\beta_2^2 r_s^2 \right)}{\left(1+\beta_1 r_s^{1/2}+\beta_2 r_s \right)^5} \right]n^{-3}, \quad r_s \geq 1.
\end{cases}
\end{eqnarray*}
\section*{Supplementary Note 2: Approximating experimental data for different $\lambda$}\label{supp:OFx}
\begin{figure}[H]
	\centering
	\includegraphics[width=0.6\linewidth]{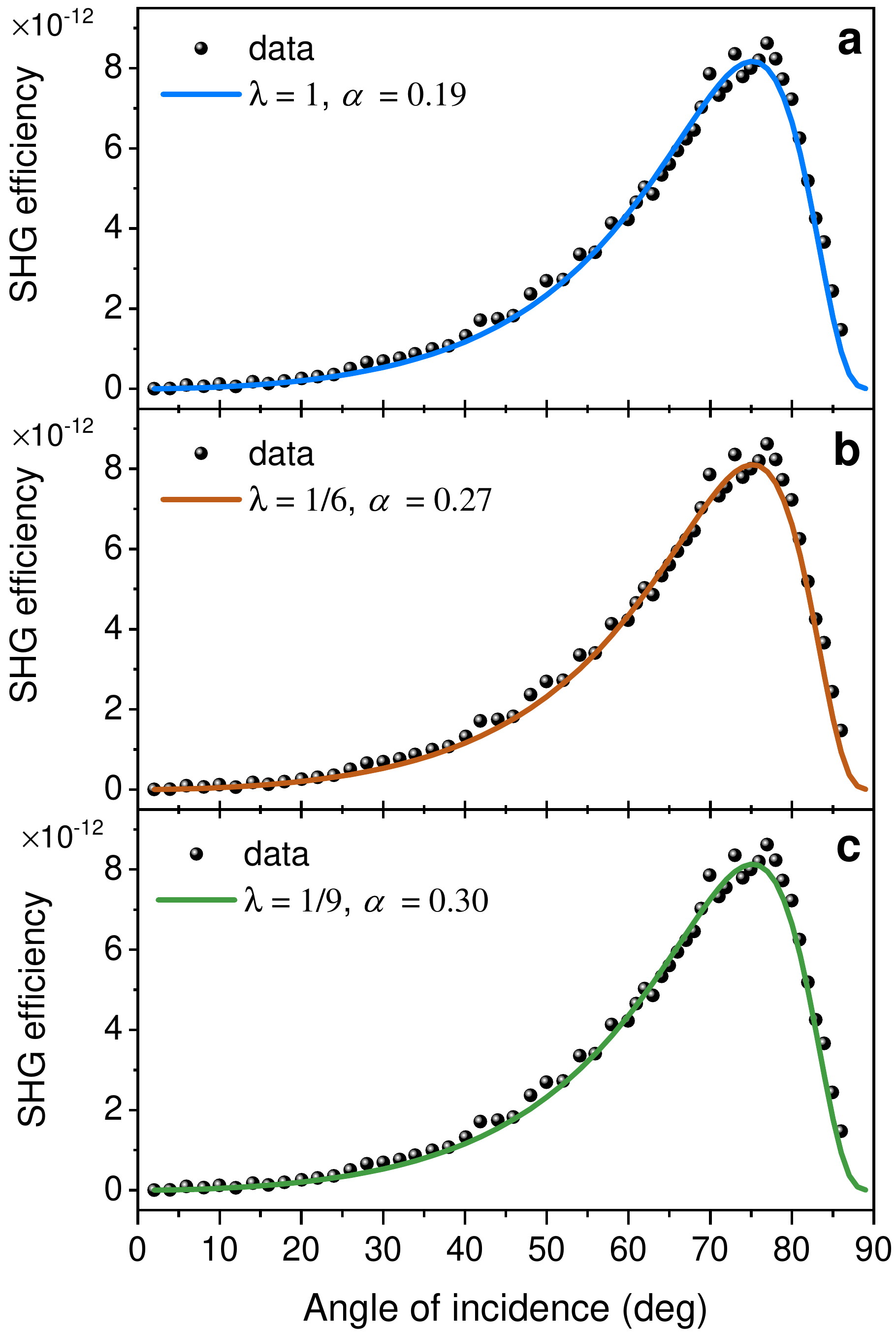}
	\caption{Nonlinear quantum hydrodynamic theory approximating the experimental data \cite{ODonnell2005NewJ.Phys.} for a 400 nm thick clean Ag film excited by a TM-polarized plane wave at $E=1.1653$~eV with a peak intensity $I_0=80$~MW/cm$^2$, considering different values of the spill-out parameter $\lambda$ and the parameter $\alpha$ weighting the nonlinear part of the von-Weizs\"acker energy functional. SHG efficiency compared against the experimental data for (\textbf{a}) $\lambda=1$ and $\alpha=0.19$, (\textbf{b}) $\lambda=1/6$ and $\alpha=0.27$, and (\textbf{c}) $\lambda=1/9$ and $\alpha=0.30$.} 
	\label{fig:Exp_etaFits}
\end{figure}

\noindent The nonlinear quantum hydrodynamic theory (QHT) can also approximate the experimental data \cite{ODonnell2005NewJ.Phys.} for Ag film for a given $\lambda$, representing the electron spill-out from the metal surface, provided that the value of the parameter $\alpha$, weighting the nonlinear part of the von-Weizs\"acker energy functional (see equation~(3) in the main text), is properly chosen. As we discussed in the main text, since the QHT is based on the approximate energy functionals, the full self-consistent approach does not reproduce well the experimental data. This is because the QHT shows a strong dependence on the equilibrium charge density which is affected by the choice of $\lambda$ and this in turn affects the SHG efficiency. The parameter $\alpha$, on the other hand, changes the relative weight of the nonlinear contributions emerging from the von Weizs\"acker terms. Therefore, there is more than one combination of $\lambda$ and $\alpha$ that enables us to fit the experimental data. Supplementary Figure~\ref{fig:Exp_etaFits} plots the SHG efficiency for a clean Ag film computed within the nonlinear QHT for different values of $\lambda$ and compared against the experimental results.

\section*{Supplementary Note 3: SHG efficiency under different pumping-detection conditions}
The data for the SHG efficiency from the Ag film in Supplementary Ref. \citenum{ODonnell2005NewJ.Phys.} was measured for three different incident-detection polarization conditions: (i) TM-incident and TM-detected; (ii) TE-incident and TM-detected, and (iii) $45\degree$ TE/TM-incident and TE-detected. Case (i) has already been reported in the main text (Fig. 3b). Here we attempt to fit the experimental data for the other two cases within the framework of QHT. A comparison of the SHG efficiency between the measured and simulated data is shown in Supplementary Figure \ref{fig:Exp_CrossPol}. The choice of the parameters is the same as used in Fig.~3b of the main text. In case (iii), we used an input intensity~$I_0=160~\rm{MW/cm^2}$. It is interesting to note that even under these polarization conditions, the QHT approximates the experimental data quite well.
\begin{figure}[H]
	\centering
	\includegraphics[width=0.6\linewidth]{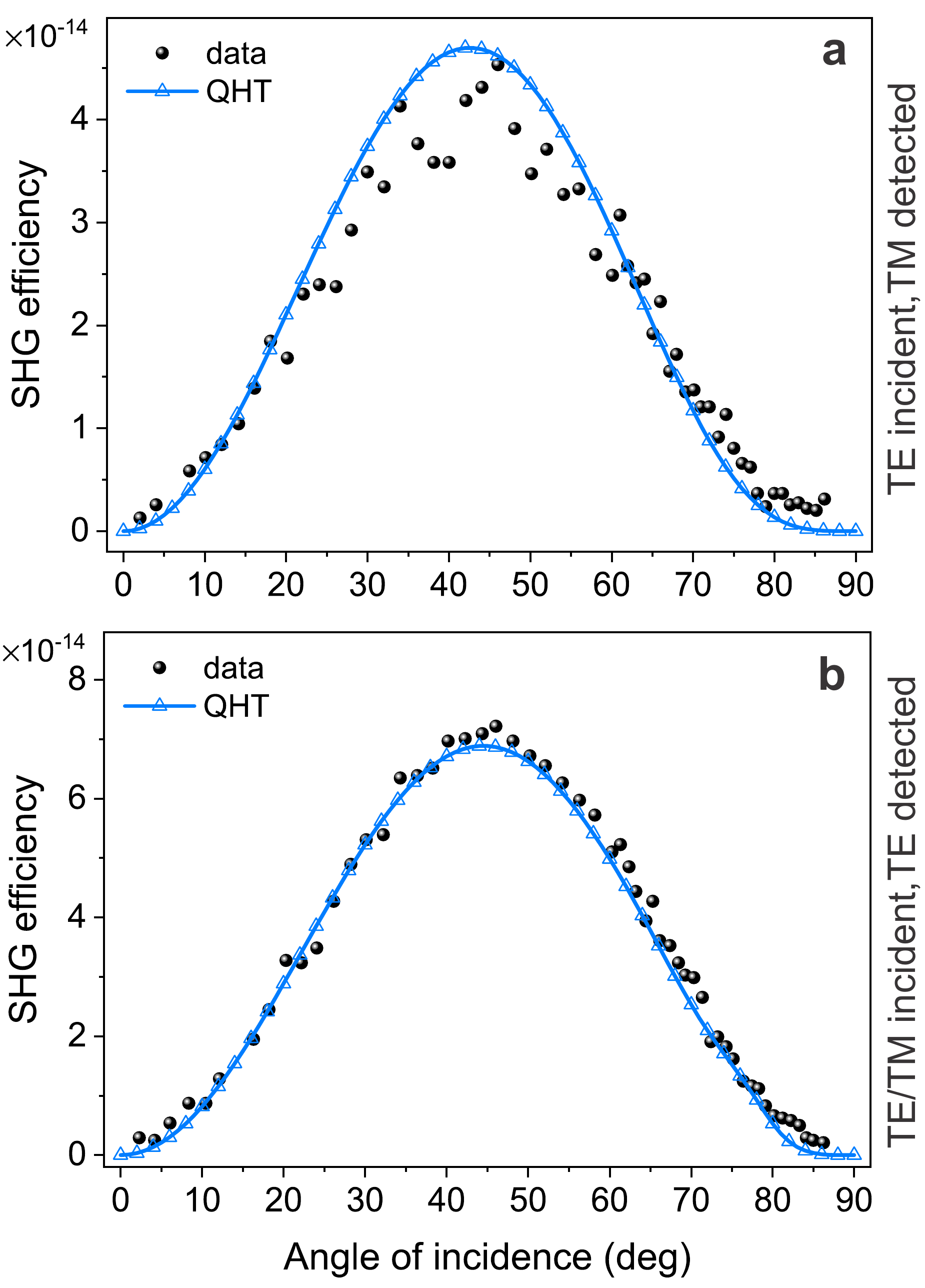}
	\caption{Experimental data \cite{ODonnell2005NewJ.Phys.} for the SHG efficiency fitted within the framework of QHT for a 400 nm thick clean Ag film under different excitation/detection polarization conditions. (\textbf{a}) TE-incident and TM-detected with an input peak intensity~$I_0=80~\rm{MW/cm^2}$, and (\textbf{b}) TE/TM-incident and TE-detected with an input intensity~$I_0=160~\rm{MW/cm^2}$. All the other parameters are the same as used in Fig.~3b of the main text.}
	\label{fig:Exp_CrossPol}
\end{figure}

\section*{Supplementary Note 4: Equilibrium charge density for various coating materials}\label{supp:n0}
\begin{figure}[H]
	\centering
	\includegraphics[width=0.58\linewidth]{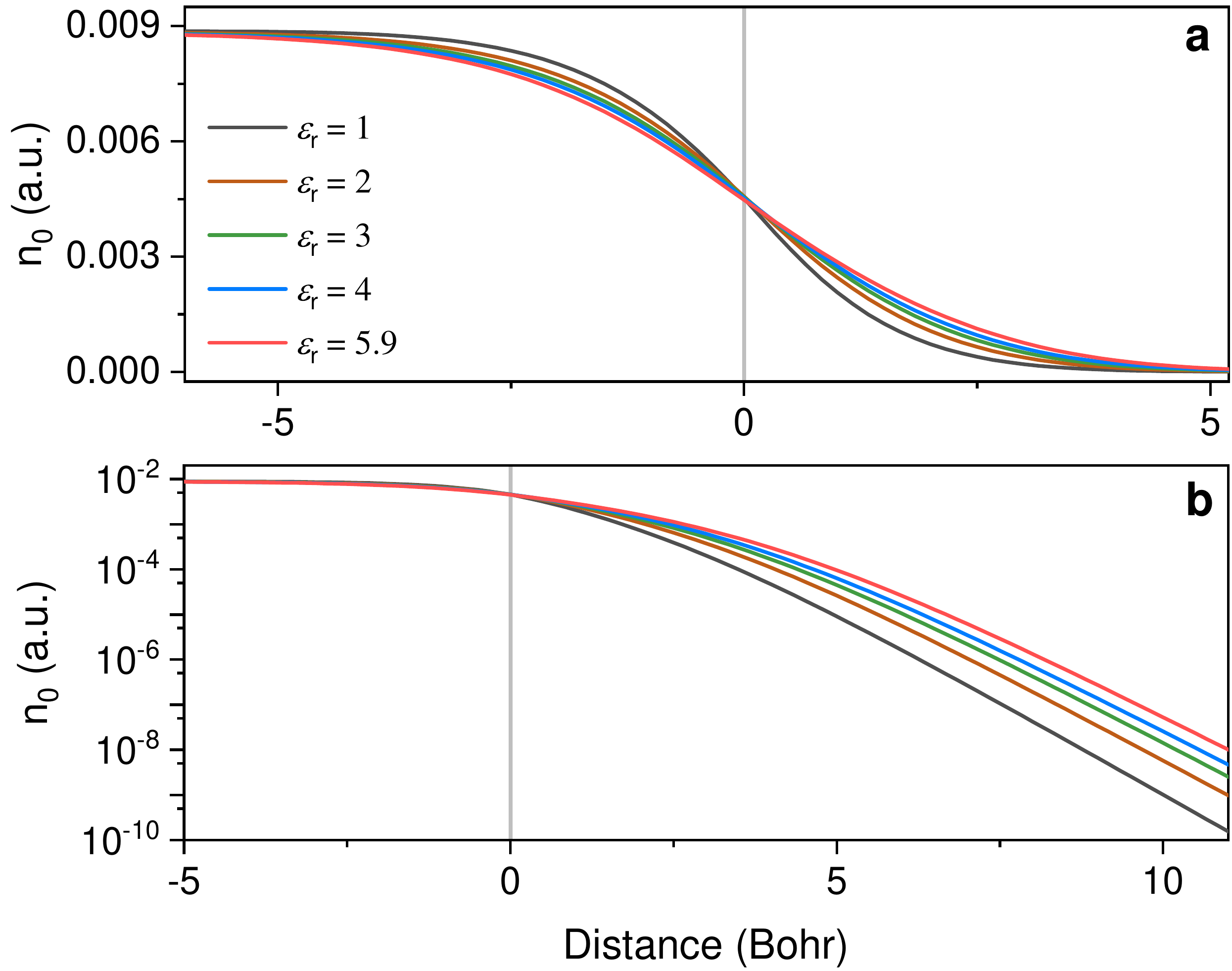}
	\caption{(\textbf{a}) Equilibrium electron density for the Ag film using the spill-out parameter $\lambda=1/4$, and considering various coating materials with a dielectric constant $\varepsilon_r$. (\textbf{b}) The decay of the charge density for different $\varepsilon_r$. The charge density for the clean Ag film, i.e., without any coating $(\varepsilon_r=1)$ is also shown. The solid grey line indicates the metal-dielectric interface.} 
	\label{fig:n0_eta4_epsr}
\end{figure}

\noindent
Supplementary Figure~\ref{fig:n0_eta4_epsr} shows the impact of the dielectric constant of the coating material on the equilibrium charge density.
It can be observed how the electron density is more loosely localized (slower decay) at the metal surface when increasing values of the dielectric constant of the coating material are considered.
This corresponds to lowering the work function $W$ (see Eq.~(4) in the main text).
\section*{Supplementary Note 5: Weak formulation}\label{supp:WeakForm}
\noindent In finite-element-method based approaches, the differential equations are usually expressed in their weak form to relax the continuity requirements for the test functions. We numerically implement equations (11)--(14) of the main text and two frequency-dependent wave equations in finite-element-method based solver \textsc{comsol} Multiphysics \cite{comsol}. In the following, we derive the weak formulation of these equations. Rewriting equations~(11) and (12) of the main text in a compact form,
\begin{equation}
-\frac{en_0}{m_e} \nabla \left(\frac{\delta G[n]}{\delta n} \right)_j-\left(\omega_j^2+i \gamma\omega_j \right) \boldsymbol{\mathrm{P}}_j=\varepsilon_0 \omega_p^2\boldsymbol{\mathrm{E}}_j+\boldsymbol{\mathrm{S}}_j, \qquad \text{where,}\quad j=1,2
\label{eq:compact}
\end{equation}
where $\boldsymbol{\mathrm{S}}_1=0$ and $\boldsymbol{\mathrm{S}}_2=\boldsymbol{\mathrm{S}}_{\rm NL}$, given by equation~(13) of the main text, as: 
\begin{equation}
\begin{split}
\boldsymbol{\mathrm{S}}_{\mathrm{NL}}=&\frac{e^2n_1}{m_e}  \boldsymbol{\mathrm{E}}_1 +i\omega_1 \frac{\mu_0 e}{m_e}\boldsymbol{\mathrm{P}}_1\times \boldsymbol{\mathrm{H}}_1 - \frac{{\omega _1^2}}{{e{n_0}}}\left( {{{\bf{P}}_1}\nabla  \cdot {{\bf{P}}_1} + {{\bf{P}}_1} \cdot \nabla {{\bf{P}}_1} - \frac{{\bf{P}}_1\cdot\nabla n_0}{n_0}{\bf{P}}_1} \right)+ \\
&+ \frac{en_1}{m_e}\nabla \left(\frac{\delta G[n]}{\delta n} \right)_1+ \frac{en_0}{m_e}\nabla \left(\frac{\delta G[n]}{\delta n} \right)^{\mathrm{NL}}_2.
\end{split}
\end{equation}
The weak formulation of Supplementary Equation~(\ref{eq:compact}) can be obtained by multiplying it with a test function $\tilde{\boldsymbol{\mathrm{P}}}_j$ and integrating over the simulation domain.
\begin{equation}
\int \left[\frac{e}{m_e} \nabla \left(\frac{\delta G[n]}{\delta n} \right)_j+\frac{1}{n_0}\left\{\left(\omega_j^2+i \gamma\omega_j \right) \boldsymbol{\mathrm{P}}_j+\varepsilon_0 \omega_p^2\boldsymbol{\mathrm{E}}_j\right\}+\frac{1}{n_0}\boldsymbol{\mathrm{S}}_j\right]\cdot\tilde{\boldsymbol{\mathrm{P}}}_j ~dV=0,
\end{equation}
\begin{equation}
\int -\frac{e}{m_e} \left(\frac{\delta G[n]}{\delta n} \right)_j \left(\nabla \cdot \tilde{\boldsymbol{\mathrm{P}}}_j\right)+\frac{1}{n_0}\left[\left(\omega_j^2+i \gamma\omega_j \right) \boldsymbol{\mathrm{P}}_j+\varepsilon_0 \omega_p^2\boldsymbol{\mathrm{E}}_j\right]\cdot \tilde{\boldsymbol{\mathrm{P}}}_j+\frac{1}{n_0}\boldsymbol{\mathrm{S}}_j\cdot\tilde{\boldsymbol{\mathrm{P}}}_j ~dV=0.
\label{eq:weakSH}
\end{equation}
Since $\boldsymbol{\mathrm{S}}_1=0$, the integral of the nonlinear source (last) term in the Supplementary Equation~(\ref{eq:weakSH}) can be expressed as:
\begin{equation}
\begin{split}
\int \frac{1}{n_0}\boldsymbol{\mathrm{S}}_{\rm NL}\cdot\tilde{\boldsymbol{\mathrm{P}}}_2 ~dV=&\int \frac{1}{n_0}\left[ \frac{e^2n_1}{m_e}  \boldsymbol{\mathrm{E}}_1 +i\omega_1 \frac{\mu_0 e}{m_e}\boldsymbol{\mathrm{P}}_1\times \boldsymbol{\mathrm{H}}_1 + \right. \\
& \left. - \frac{{\omega _1^2}}{{e{n_0}}}\left( {{{\bf{P}}_1}\nabla  \cdot {{\bf{P}}_1} + {{\bf{P}}_1} \cdot \nabla {{\bf{P}}_1} - \frac{{\bf{P}}_1\cdot\nabla n_0}{n_0}{\bf{P}}_1} \right)\right]\cdot \tilde{\boldsymbol{\mathrm{P}}}_2+ \\
&+ \frac{en_1}{m_en_0}\nabla \left(\frac{\delta G[n]}{\delta n} \right)_1\cdot \tilde{\boldsymbol{\mathrm{P}}}_2+ \frac{e}{m_e}\nabla \left(\frac{\delta G[n]}{\delta n} \right)^{\mathrm{NL}}_2\cdot \tilde{\boldsymbol{\mathrm{P}}}_2 ~dV.
\end{split}
\label{eq:S_NL}
\end{equation}
The last two terms appearing in the above equation emerge from the spatial dependence of the charge density. Thus, its weak form can be finally written as: 
\begin{equation}
\begin{split}
\int \frac{1}{n_0}\boldsymbol{\mathrm{S}}_{\rm NL}\cdot\tilde{\boldsymbol{\mathrm{P}}}_2 ~dV=&\int \frac{1}{n_0}\left[ \frac{e^2n_1}{m_e}  \boldsymbol{\mathrm{E}}_1 +i\omega_1 \frac{\mu_0 e}{m_e}\boldsymbol{\mathrm{P}}_1\times \boldsymbol{\mathrm{H}}_1 + \right. 
\\
& \left. - \frac{{\omega _1^2}}{{e{n_0}}}\left( {{{\bf{P}}_1}\nabla  \cdot {{\bf{P}}_1} + {{\bf{P}}_1} \cdot \nabla {{\bf{P}}_1} - \frac{{\bf{P}}_1\cdot\nabla n_0}{n_0}{\bf{P}}_1} \right)\right]\cdot \tilde{\boldsymbol{\mathrm{P}}}_2+ 
\\
& - \frac{e}{m_e} \left(\frac{\delta G[n]}{\delta n} \right)_1 \left[\frac{n_1}{n_0} \left(\nabla \cdot \tilde{\boldsymbol{\mathrm{P}}}_2\right)+\frac{1}{n_0^2} \left(n_0 \nabla n_1-n_1\nabla n_0 \right)\cdot \tilde{\boldsymbol{\mathrm{P}}}_2 \right]+ \\
& - \frac{e}{m_e} \left(\frac{\delta G[n]}{\delta n} \right)^{\mathrm{NL}}_2 \left(\nabla \cdot \tilde{\boldsymbol{\mathrm{P}}}_2 \right) ~dV.
\end{split}
\label{eq:S_NL_Weak}
\end{equation}
The advantage of the weak formulation is that there is no need to compute the gradients of the energy functionals, and the derivatives are distributed to the test functions. The energy functionals also contain second-order derivatives, and we introduced working variables, such that, $\boldsymbol{\mathrm{F}}_j=\nabla n_j$ where $n_j=\frac{1}{e}\nabla \cdot \boldsymbol{\mathrm{P}}_j$ which implies that $\nabla \cdot \boldsymbol{\mathrm{F}}_j=\nabla^2 n_j$.
Thus, the complete set of nonlinear QHT equations, along with the two wave equations and a static QHT equation for computing equilibrium density, in their weak formulation reads,
\begin{eqnarray}
& &\int \left(\nabla \times \boldsymbol{\mathrm{E}}_j \right) \cdot \left(\nabla \times \tilde{\boldsymbol{\mathrm{E}}}_j \right)-\left( \varepsilon_\infty \frac{\omega_j^2}{c^2}\boldsymbol{\mathrm{E}}_j+\mu_0 \omega_j^2 \boldsymbol{\mathrm{P}}_j \right)\cdot \tilde{\boldsymbol{\mathrm{E}}}_j ~dV=0, \\
& &\int -\frac{e}{m_e} \left(\frac{\delta G[n]}{\delta n} \right)_j \left(\nabla \cdot \tilde{\boldsymbol{\mathrm{P}}}_j\right)+\frac{1}{n_0}\left[\left(\omega_j^2+i \gamma\omega_j \right) \boldsymbol{\mathrm{P}}_j+\varepsilon_0 \omega_p^2\boldsymbol{\mathrm{E}}_j\right]\cdot \tilde{\boldsymbol{\mathrm{P}}}_j+\frac{1}{n_0}\boldsymbol{\mathrm{S}}_j\cdot\tilde{\boldsymbol{\mathrm{P}}}_j ~dV=0, \\
& &\int -\left(\nabla \cdot \boldsymbol{\mathrm{P}}_j \right)\left(\nabla \cdot \tilde{\boldsymbol{\mathrm{F}}}_j \right)-e \boldsymbol{\mathrm{F}}_j \cdot \tilde{\boldsymbol{\mathrm{F}}}_j ~dV=0, \\
& &\int - \varepsilon_\infty(\boldsymbol{\rm r})\nabla \left(\frac{\delta G[\xi^2]}{\delta n} \right)_{\xi=\sqrt{n_0}} \cdot \nabla \tilde{\xi}+\frac{e^2}{\varepsilon_0}\left(\xi^2-n^+ \right)\tilde{\xi} ~dV=0, \label{eq:n0_weak}
\end{eqnarray} 
where $j=1,2$ and $\boldsymbol{\mathrm{S}}_1=0$. In Supplementary Equation~~(\ref{eq:n0_weak}), the transformed variable $\xi=\sqrt{n_0}$ was introduced as in this way the solution converges more quickly.
\def\bibsection{\section*{Supplementary References}}
\bibliography{myLibrary}